\pdfoutput=1
\documentclass[aps,prl,twocolumn,noshowpacs,a4paper,superscriptaddress]{revtex4-1}

\usepackage{graphicx}
\usepackage{color}
\usepackage{textcomp}
\usepackage{dsfont}
\usepackage{amssymb}\usepackage{amsmath}\usepackage{amsthm}
\usepackage{units}

\usepackage[normalem]{ulem}

\begin{document}
\title{Phase Transitions in a Bose-Hubbard Model with Cavity-Mediated Global-Range Interactions}

\author{N.~Dogra}
\affiliation{Institute for Quantum Electronics, ETH Z\"{u}rich,
8093 Z\"{u}rich, Switzerland}

\author{F.~Brennecke}
\affiliation{Physikalisches Institut, University of Bonn, Wegelerstrasse 8, 53115 Bonn, Germany}

\author{S.D.~Huber}
\affiliation{Institute for Theoretical Physics, ETH Z\"{u}rich,
8093 Z\"{u}rich, Switzerland}

\author{T.~Donner}
\affiliation{Institute for Quantum Electronics, ETH Z\"{u}rich,
8093 Z\"{u}rich, Switzerland}

\date{\today}

\begin{abstract}
We study a system with competing short- and global-range interactions in the framework
of the Bose-Hubbard model. Using a mean-field approximation we obtain the phase diagram of the system and observe four different phases: a superfluid, a supersolid, a Mott insulator and a charge density wave, where the transitions between the various phases can be either of first or second order.  We qualitatively support these results using Monte-Carlo simulations. An analysis of the low-energy excitations shows that the second-order phase transition from the charge density wave to the supersolid is associated with  the softening of particle- and hole-like excitations which give rise to a gapless mode and an amplitude Higgs mode in the supersolid phase. This amplitude Higgs mode is further transformed into a roton mode which softens at the supersolid to superfluid phase transition.
\end{abstract}
\maketitle

Ultracold atomic gases confined in optical lattices have proven to be impressively successful in simulating strongly correlated lattice models like the Bose-Hubbard model, which features a quantum phase transition from a superfluid to a Mott insulating phase \cite{Jaksch1998a, Greiner2002}. Extended Bose-Hubbard models hosting long-range interactions have been theoretically predicted to show additional intriguing quantum phases such as supersolids or charge density waves when short-range interactions and kinetic energy compete with long-range interactions. However, their experimental demonstration has so far been elusive despite strong efforts in the creation of quantum gases with long-range dipolar interactions \cite{Baier2016, Yan2013}. Only recently, a complimentary approach using cavity-mediated global-range interactions \cite{Mottl2012a} in combination with static optical lattices \cite{Klinder2015a} could reach the regime where short- and long-range interactions compete \cite{Landig2016} which led to the observation of a rich phase diagram with Mott insulating, superfluid, supersolid and charge density wave phases.

Gaining deeper understanding of the emerging quantum phases and phase transitions in such a system requires the study of an extended Bose-Hubbard model incorporating global-range interactions with an underlying $\mathds{Z}_2 $-symmetry \cite{Ritsch2013}. We consider a situation conceptually similar to the experiment \cite{Landig2016} where ultracold bosons in an optical lattice potential are strongly coupled to a single mode of a high-finesse optical cavity and illuminated with a transverse standing wave laser field. Photon scattering from the laser field into the cavity mode and back induces global-range interactions \cite{Mottl2012a} . Short-range interactions arise due to the s-wave scattering between atoms residing on the same site of the optical lattice. These two energy scales can compete with each other and with the kinetic energy of the particles. Previous theoretical work on the combination of Bose-Hubbard models with dynamical cavity-fields \cite{Larson2008a, Fernandez-Vidal2010, Ritsch2013,  Habibian2013, Habibian2013a, Li2013, Bakhtiari2015, Caballero-Benitez2015} discussed some of the resulting phases and, very recently, predicted the phase diagram \cite{Chen2016a}. In this work, we use a mean-field ansatz to derive the phase diagram both as a function of chemical potential and -- experimentally more relevant -- as a function of densities, and analyze the type of the involved phase transitions. We support this finding with a quantum Monte Carlo simulation. Further, employing a  slave-Boson approach we get access to the low-lying excitations of the system which we can identify as particle- and hole-like excitations, a sound mode and an amplitude Higgs mode that transforms into a roton-like excitation.

\begin{figure*}
\includegraphics[width=1.75\columnwidth]{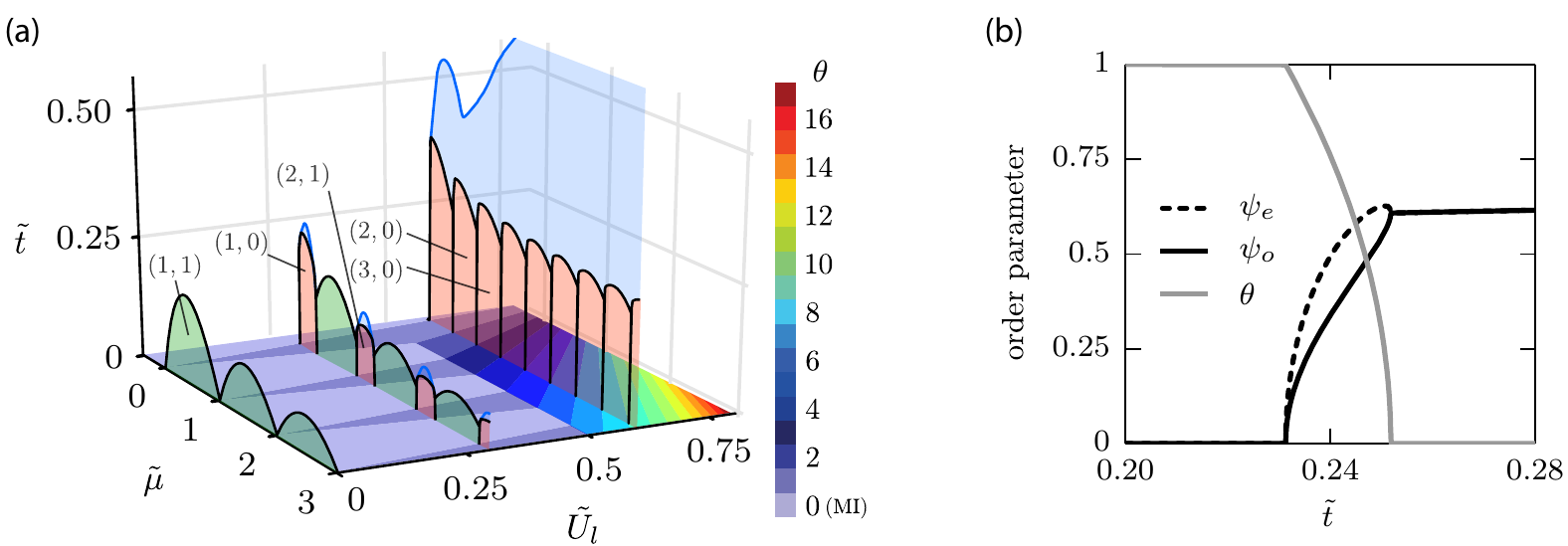}
\caption{(a) Phase diagram as function of rescaled parameters $\tilde{\mu}$, $\tilde{t}$ and $\tilde{U_l}$. The colorbar refers to $\tilde{t}=0$ and indicates the imbalance $\theta$ between the number of atoms on even and odd sites. The transparent colors show the different phases for $\tilde{t}>0$: MI (green), CDW (red), and SS (blue). The SF phase is not indicated, but fills the remaining space.  The labels correspond to the site populations ($n_e,n_o)$. (b) Evolution of the different order parameters as a function of tunneling $\tilde{t}$ for filling $\rho = 1/2$, corresponding to a cut through the CDW(1,0)-lobe at $\tilde{U_l}=0.3$ in panel (a). With increasing $\tilde{t}$, the system evolves from CDW over SS to SF. Here, all order parameters change continuously across the phase transitions signaling that these are of second order.} 
\label{fig1}
\end{figure*}

We consider an ultracold gas of bosonic atoms tightly confined to the $x$-$z$ plane, and subject to a square optical lattice with lattice constant $a=\lambda/2$ in the $x$- and $z$-directions. Here, $\lambda$ is the wavelength of the laser field creating the lattices. The atomic gas is strongly coupled to a fundamental mode of a high-finesse Fabry-Perot cavity oriented along the $x$-direction. The frequency $\omega_z$ of the $z$-lattice laser field, which is polarized along the $y$-direction, is detuned with respect to the dispersively shifted cavity resonance frequency $\omega_c$ by $\Delta_c = \omega_z - \omega_c$. The coherent scattering of light between the $z$-lattice and the cavity mode at two-photon Rabi frequency $\eta$ creates a dynamical checkerboard superlattice potential of periodicity $\lambda$ for the atoms \cite{Baumann2010}, corresponding effectively to global-range atom-atom interactions \cite{Mottl2012a}. For a many-body description of the system we expand the bosonic field operators for the atoms in terms of Wannier functions $w(\mathbf{r}-\mathbf{r}_i)$ localized at lattice site $i$ of the externally imposed square optical lattice potential \cite{Maschler2008a}. We further assume the cavity decay rate $\kappa$ and the detuning $\Delta_c$ to be large compared to the inverse time scale of atomic motion which allows us to adiabatically eliminate the cavity field dynamics from the equations of motion \footnote{For the experimental parameters of \cite{Landig2016}, we estimate the relative shift in excitation energies of the atomic branch of the polariton due to this adiabatic elimination to be on the order of $10^{-4}$ and thus negligible.}. The effective Hamiltonian describing the atomic dynamics, dressed by the cavity field, of a uniform system with chemical potential $\mu$ is then given by

\begin{equation}
\begin{split}
\hat{H} =  - t \sum_{\langle i,j \rangle}\left(\hat{b}_i^\dag \hat{b}_{j} + \mathrm{h}.\mathrm{c}.\right) + \frac{U_s}{2} \sum_{i\in e,o} \hat{n}_{i}(\hat{n}_{i}-1) \\
   - \frac{U_l}{K}\left(\sum_{i\in e}{\hat{n}_{i}} - \sum_{i\in o}{\hat{n}_{i}}\right)^{2} - \mu\sum_{i\in e,o}{\hat{n}_{i}}\,.
\end{split}
\label{BHHam}
\end{equation}

Here, $\hat{b}^\dag_{i}$ ($\hat{b}_{i}$) defines the atomic creation (annihilation) operator and $\hat{n}_i = \hat{b}^\dag_{i}\hat{b}_{i}$ is the corresponding number operator. The symbol $\langle i,j \rangle$ indicates that the summation runs over all pairs of sites. The subscript $e$($o$) refers to even (odd)  lattice sites $i=(i_x,i_z)$ of the square lattice potential defined as $i_x+i_z \in $ even (odd). The first two terms in Eq. \eqref{BHHam} represent the  hopping of particles with amplitude $t$ between adjacent sites $i,j$, and the onsite interaction between the particles with strength $U_s$ \cite{Jaksch1998a}. The third term describes the inter-particle global-range interaction with strength $U_l$ given by
\[
U_l = -K\hbar M_0^2\eta^2\frac{\Delta_c}{\Delta_c^2 + \kappa^2} \stackrel{\text{$|\Delta_{\rm{c}}| \gg \kappa$}}{\approx} -K\hbar M_0^2\frac{\eta^2}{\Delta_c} \propto \frac{V}{\Delta_c}
\]
where $K$ is the total number of lattice sites. $M_0$ is the overlap between the Wannier function $w(\mathbf{r})$ and the checkerboard potential $\propto \cos{\frac{2\pi x}{\lambda}} \cos{\frac{2\pi z}{\lambda}}$, which is induced by the interference of the z-lattice laser field with the light field scattered into the cavity. We neglect dynamics of the dispersive shift of the cavity resonance due to density redistributions of the atoms as well as the influence of the additional lattice potential induced by the field scattered into the cavity (see Appendix B). The relative strength of global-range interactions $U_l$, short range interactions $U_s$ and tunneling $t$ can be independently tuned by changing the lattice depth $V$ and the detuning $\Delta_c$ which is assumed in the following to be negative such that $U_l > 0$.

As we show in Appendix A, Hamiltonian (\ref{BHHam}) describing the system with global-range interactions is on a mean-field level formally equivalent to the Bose-Hubbard model including nearest-neighbor interactions. This equivalence puts experimental and theoretical results on the extended Bose-Hubbard model with cavity-mediated interaction into a much broader context.

The global-range interaction term in Eq. \ref{BHHam} is proportional to the square of the total  imbalance $\hat{\Theta} = \sum_{i\in e}\hat{n}_i-\sum_{i\in o}\hat{n}_i$ between the number of atoms populating even and odd sites. The independence of this term on the sign of the imbalance reflects the underlying $\mathds{Z}_2$-symmetry. This term thus just favors an overall even-odd imbalance. Any additional local structure of the atomic density distribution is dictated by the relative strength of the short-range interaction and the tunneling.

To determine the ground state phase diagram, we define three order parameters, $\psi_{e/o} = \langle \hat{b}_{e/o}\rangle$ which are the superfluid order parameters for the even and odd lattice sites, and $\theta = 2\langle \hat{\Theta} \rangle/K$ which characterizes the average imbalance between the two kinds of sites. We obtain an approximate expression for these order parameters, which we choose without  loss of generality real-valued, by employing a mean-field decoupling of the kinetic energy and the global-range interaction  \cite{VanOosten2001}:

\begin{equation*}
  \hat{b}_{e}^{\dagger}\hat{b}_{o} + \hat{b}_{o}^{\dagger}\hat{b}_{e} \approx \psi_e (\hat{b}_{o} + \hat{b}_{o}^{\dagger})+ \psi_o(\hat{b}_{e}^{\dagger} + \hat{b}_{e}) - 2\psi_e\psi_o
 \end{equation*}
 
 \begin{equation}
    \hat{\Theta}^{2} \approx 2\langle \hat{\Theta} \rangle \hat{\Theta} - \langle \hat{\Theta} \rangle ^{2} = K\theta\hat{\Theta} - \frac{K^2}{4}\theta^2 \,.  \label{ThetaApprox} 
 \end{equation}

This approach reduces the description to an effective two-site problem where the different order parameters are calculated self-consistently with the ground state of the system using an iterative algorithm \cite{Dhar2011}. In the experiment, the $\mathds{Z}_2$-symmetry will be spontaneously broken and we choose without loss of generality $\theta \ge 0$.

We display in Fig.~\ref{fig1}a the phase diagram as a function of chemical potential $\tilde{\mu}=\mu/U_s$, tunneling strength $\tilde{t}=z t/U_s$ and global-range interaction strength $\tilde{U_l}=U_l/U_s$, each normalized with respect to the short-range interaction strength $U_s$. The coordination number is labeled $z$. In the zero-tunneling limit (bottom plane in Fig.~\ref{fig1}a), we find besides the Mott insulator (MI) states additional charge density wave (CDW) states with a finite even-odd imbalance $\theta$. For $0 < \tilde{U_l} < 0.5$, only CDW states with an imbalance of $\theta = 1$ are found, which lie in a chemical potential range of
\[
n_e-1-\frac{\tilde{U_l}}{2}(2\theta-1)< \tilde{\mu} < n_o + \frac{\tilde{U_l}}{2} (2\theta - 1)\,.
\]
Here,  $n_{e,o} = \langle \hat{n}_{e,o}\rangle$, and $\theta$ is 0(1) for MI(CDW) state. Beyond  $\tilde{U_l}=0.5$, the topology of the phase diagram changes \cite{Caballero-Benitez2015}, and one finds only CDW states with maximum imbalance ($\theta = \langle \hat{n}_e \rangle$, $\langle \hat{n}_o \rangle = 0$). In this limit, the chemical potential range of the CDW states is
\[
\theta-1-\frac{\tilde{U_l}}{2}\big(2\theta-1\big) < \tilde{\mu} < \theta - \frac{\tilde{U_l}}{2}\big(2\theta + 1\big)\,.
\]

For a non-zero tunneling amplitude, a supersolid (SS) phase, characterized by the presence of a finite even-odd imbalance and non-zero superfluid order parameters, is observed at the tip of the CDW phase (see blue regions in Figure~\ref{fig1}(a)). Upon further increasing tunneling, we obtain a superfluid (SF) phase with finite and equal superfluid order parameters and vanishing even-odd imbalance. An exemplary evolution of the order parameters across the different phases is shown in Figure~\ref{fig1}(b) for $\tilde{U_l} = 0.3$ and a constant filling $\rho = 1/2$, where we define the filling as the average number of particles per lattice site of the underlying square lattice.

Increasing the strength of global-range interactions increases the region of CDW and SS phases both as a function of chemical potential and tunneling strength, see Figure~\ref{fig1}a. For $\tilde{U}_l>0$, the system preserves the insulating character for finite but small tunneling for all chemical potentials. For $\tilde{U_l} > 0.5$, no MI exists anymore, and the SS region is observed in a large connected parameter regime. Since Hamiltonian (\ref{BHHam}) is formally equivalent to a system with nearest-neighbor interactions, this result is in line with a similar prediction for the case of dipolar gases \cite{Iskin2011}. The description of the system in terms of the grand canonical ensemble becomes invalid for $\tilde{U_l} \geq 1$ because the global-range interaction term which scales quadratically with the number operators becomes strong enough to generate an arbitrarily large imbalance. This in turn leads to the divergence of the filling, independent of the chemical potential.  

\begin{figure}[h]
\includegraphics[width=1\columnwidth]{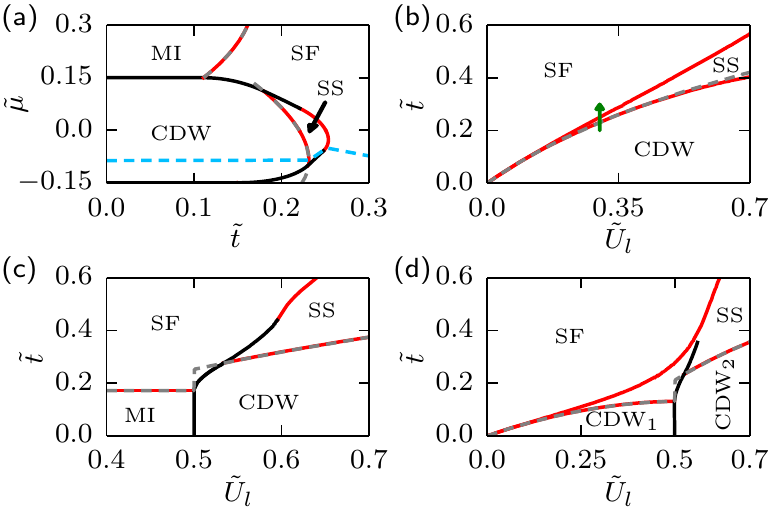}
\caption{Red (black) lines indicate second-(first-) order transitions. The dashed grey lines represent  phase boundaries found using second-order perturbation theory.  (a) Phase diagram as function of $\tilde{\mu}$ and $\tilde{t}$ for $\tilde{U_l}=0.3$. The dashed blue line is the chemical potential corresponding to a fixed filling $\rho = 1/2$. (b-d) Phase diagram as a function of $\tilde{U_l}$ and $\tilde{t}$ for fixed fillings of $\rho = 1/2$ (b), $\rho = 1$ (c), and $\rho = 3/2$ (d).  The green arrow in (b) indicates the parameter range which is used to study the second-order phase transition from CDW to SS and eventually to a SF phase in Figure~\ref{fig4}, and also the evolution of the order parameters in Figure~\ref{fig1}(b). $\mathrm{CDW_1}$ and $\mathrm{CDW_2}$ correspond to a filling of $(n_e,n_o)=(2,1)$ and $(n_e,n_o)=(3,0)$, respectively.}
\label{fig2}
\end{figure}

We classify the phase transitions in Figure~\ref{fig1}a via the change in the various order parameters across the phase boundary, where a transition is of second (first) order if this change is (dis-)continuous. This is shown in Figure~\ref{fig2}(a) for $\tilde{U_l} = 0.3$, where the red lines correspond to second-order phase transitions from a CDW(SS) to a SS(SF) state. The boundary for the second-order phase transition from a CDW phase to a SS phase can also be obtained analytically from second-order perturbation theory \cite{VanOosten2001}:
\begin{equation}
\tilde{t}\,^2 = \frac{1}{f(n_e,\tilde{\mu}, \tilde{U_l}\theta)f(n_o,\tilde{\mu},-\tilde{U_l}\theta)}\,.
\label{second_order}
\end{equation}
Here, $f(x,y,z) = \frac{x}{-y + x-1 -z} + \frac{x+1}{y - x + z}$ and the insulating state of the system in the zero tunneling limit is assumed to be $n_e = \langle \hat{n}_e \rangle$ and $n_o = \langle \hat{n}_o \rangle$. This phase boundary is plotted in Figure~\ref{fig2}(a) as dashed grey line and matches well with the numerical result (red). In the thermodynamic limit, there is no effect of the global-range interactions on the location of the SF to MI transition as the gain in global-range interaction energy when creating a particle- or hole-like excitation on top of a MI state is vanishingly small. This can also be seen from Eq.\eqref{second_order} as the dependence on $\tilde{U_l}$ vanishes at the SF-MI boundary where $\theta=0$.

The black lines in Figure~\ref{fig2}(a) represent first-order phase transitions. For fixed chemical potential they are accompanied by jumps in the density across the transition point which is not realistic from an experimental perspective. Therefore, we plot in Figure~\ref{fig2}(b-d) the phase diagram as a function of $\tilde{t}$ and $\tilde{U_l}$ for fixed filling which is obtained by accordingly adjusting the chemical potential. We consider the situation of only integer or half integer fillings which are the only cases where one obtains an insulating state for a homogeneous system. In Figure~\ref{fig2}(b), we display the case of $\rho = 1/2$. The CDW to SS transition boundary can also be obtained from second-order perturbation theory since this corresponds to the largest critical tunneling as a function of chemical potential in the CDW lobe for a given $\tilde{U_l}$ and is depicted by the grey dashed line. The blue dashed line in Figure~\ref{fig2}(a) indicates the corresponding variation of chemical potential across the CDW lobe for $\rho = 1/2$ and $\tilde{U_l} = 0.3$.

Figure~\ref{fig2}(c) shows the phase diagram for unity filling, $\rho = 1$. In this case, we observe a first-order phase transition from MI to CDW at $\tilde{U_l}=0.5$ and small tunneling, signaled by a jump in the imbalance order parameter. For $\tilde{U_l}<0.5$, one observes the usual MI to SF transition upon increasing the tunneling amplitude, which has no dependence on $\tilde{U_l}$. For $\tilde{U_l}\gg 0.5$, there is a second-order phase transition from CDW to SS (red line) which also matches with the results of the second-order perturbation theory (grey dashed line). For increasing tunneling a transition to a SF phase (red line) takes place, while in the vicinity of $\tilde{U_l} = 0.5$ the transitions to a SF phase are of first-order (black line). At unity filling, the system exhibits two tri-critical points where three phases meet. In Figure~\ref{fig2}(d) we finally plot the phase diagram for $\rho = 3/2$ where we observe a first-order phase transition between two CDW phases at $\tilde{U_l} = 0.5$ with a jump in the imbalance $\theta$ from 1 to 3. The first-order phase transition line extends into the SS phase where it ends in a critical point. When crossing this line, the imbalance order parameter $\theta$ shows a jump whose height decreases to zero when approching the critical point at the end of the line.

The first-order transitions at the critical value $\tilde{U_l} = 0.5$ are a general feature of this system, appearing at all integer and half-integer fillings. They take place between either a MI or a partially imbalanced CDW and a maximally imbalanced CDW state with either $n_o = 0$ or $n_e = 0$. At the critical value $\tilde{U_l} = 0.5$, the system has many  degenerate ground states with different values of $\theta$, separated from each other by large energy barriers. Away from the critical value this degeneracy gets lifted giving rise to global and local energy minima. Such an energy landscape can lead to a hysteretic behavior while changing the relative strength $\tilde{U_l}$, which is a characteristic feature of a first-order phase transition.  In the atomic limit and for $\theta>0$ and $K$ even, the positions of the energy minima are given by $\theta_{\rm{min}} = 2(\rho - \alpha)$, where $\alpha$, $\beta$ are integers and $\rho$ is the filling. For $\rho$ being integer or half integer, $0 \leq \alpha \leq \rho$, $0 \leq \beta < \rho$ with $2\rho + 1$ minima. Only these cases allow for a MI or a CDW state. For other  densities, $0 \leq \alpha,\beta < 2\rho$ and the total number of minima is given by $2\lceil 2\rho \rceil$, where $\lceil x \rceil$ refers to the smallest integer greater than $x$.

\begin{figure}[]
\includegraphics[width=1\columnwidth]{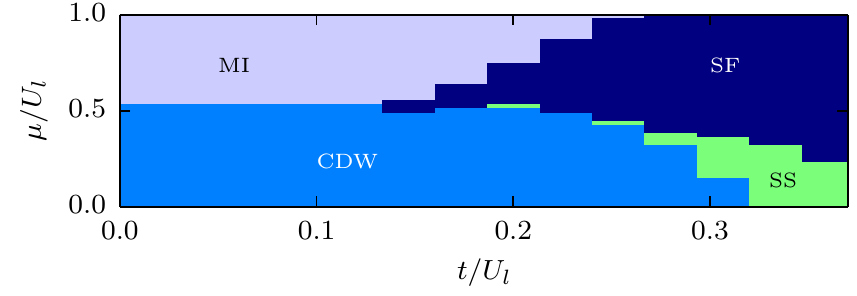}
\caption{Phase diagram resulting from finite-size scaling of a Monte-Carlo simulation in the hard-core Boson limit, i.e. $U_s=\infty$, for finite temperature $k_B T / U_l = 1/15$ as a function of chemical potential $\mu$ and tunneling $t$, both scaled with respect to $U_l$. Maximum system size is $L_x=L_y=8$. All phases and their topology found in the mean-field analysis are qualitatively reproduced.}
\label{fig3}
\end{figure}

In order to validate the phase diagram obtained in mean-field approximation, we performed quantum Monte-Carlo simulations in the limit of hard-core bosons \cite{Albuquerque2007}. We obtain the superfluid stiffness and the CDW order parameter from the simulations for system sizes of $L^2$ lattice sites with $L=4, 6, 8$, and extrapolate these values to $L=\infty$. We then check for a residual order parameter which allows us to identify the phases. The result is shown in Fig.~\ref{fig3},  qualitatively supporting the above described mean-field approach.

To further characterize the nature of the phases and corresponding transitions, an analysis of the low-energy excitations is performed. We obtain an effective Hamiltonian using a variational mean field approach \cite{Huber2007}, where we define a set of operators for the even and odd sites to generate the ground state and the low energy excitations. We limit ourselves to a three state description: $|m\rangle$, $|m-1\rangle$ and $|m+1\rangle$ for the even sites and $|n\rangle$, $|n-1\rangle$ and $|n+1\rangle$ for the odd sites where the ground state in the insulator lobe is of the form: $n_e = m$ and $n_o = n$ if $n_o \neq 0$ or $n_o = n-1$ if $n_o = 0$. The operator to generate the ground state is obtained by minimizing the free energy of the system in this three state basis for the even and odd sites using a Gutzwiller ansatz.

We obtain an effective Hamiltonian $\hat{H}_\mathrm{exc}$ describing the low-energy excitations by imposing the completeness constraint of these operators and assuming only small fluctuations on top of the ground state.  This Hamiltonian is finally diagonalized in momentum space and used to analyse the properties of the low-energy excitations.

\begin{figure}[]
\includegraphics[width=1.0\columnwidth]{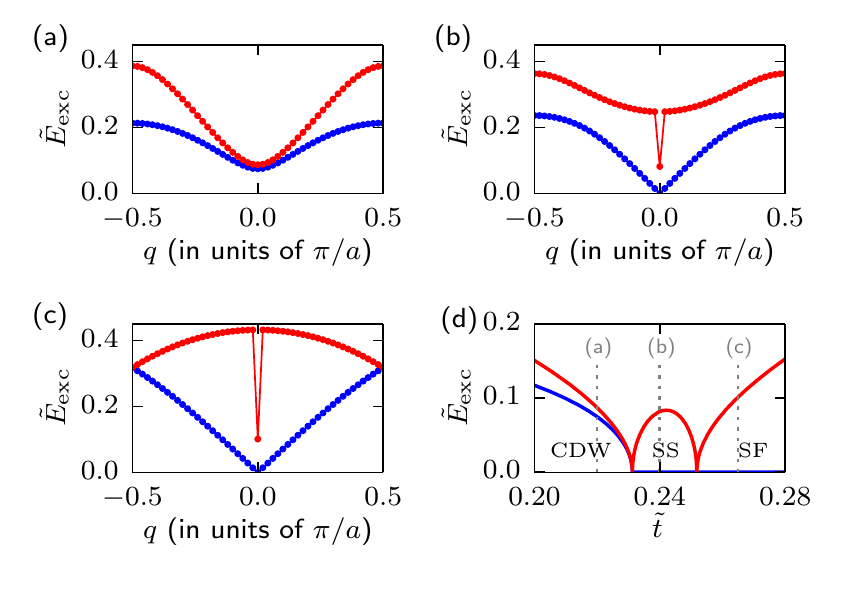}
\caption{Excitation spectra for the two lowest energy excitations in (a) CDW, (b) SS and (c) SF phase for $\tilde{U_l} = 0.3$ as a function of quasi-momentum $q$ for fixed density. (d) The two lowest excitation energies for $q=0$ as a function of tunneling. The locations of the full excitation spectra, panels (a-c), are indicated by grey dashed lines.} 
\label{fig4}
\end{figure}

Fig.~\ref{fig4} shows the two lowest excitations in the different phases for $\tilde{U_l} = 0.3$ and a fixed filling of $\rho = 1/2$. For simplicity we here consider a one-dimensional situation with $\tilde{E}_{\mathrm{exc}} = E_\mathrm{exc}/U_s$ being the excitation energy normalized with respect to the short-range interaction strength $U_s$ and $q$ being the quasi-momentum in units of $\pi/a$.  The size of the first Brillouin zone is halved due to the discrete $\mathds{Z}_2$-symmetry breaking of the underlying $\lambda/2$-periodic lattice in the CDW and SS phase and gives rise to a gap between the two lowest excitations at the boundary of the Brillouin zone in these phases, Fig.~\ref{fig4}(a),(b). Both lowest energy excitations in the CDW phase, shown in Fig.~\ref{fig4}(a), are gapped at $q=0$ as expected for an insulating phase. They correspond to a hole-like excitation (blue line) and a particle-like excitation (red line), delocalized over all even and odd sites, respectively. This is in contrast to a MI state where both the particle- and hole-like excitations are delocalized over all lattice sites.

The lowest energy excitation for the SS state, Fig.~\ref{fig4}(b), corresponds to a gapless mode of the system at $q=0$ (blue line) with a finite group velocity at small quasi-momenta. The second excitation mode (red line) shows a discrete change in the excitation energy at $q=0$ which results from the global range of the cavity-mediated interactions. In the SF phase, Fig.~\ref{fig4}(c), the lowest energy excitation is again a gapless mode (blue line) similar to the SS phase, and the second excitation (red line) is influenced by $U_l$ exclusively at $q=0$.

Due to the global range of the interactions, the most interesting evolution of the excitation energies takes place at $q=0$. We plot the energy of the two lowest excitations for $q=0$ as a function of $\tilde{t}$ in Fig.~\ref{fig4}(d), again for $\tilde{U_l} = 0.3$ and a fixed filling of $\rho=1/2$. The second-order phase transition between CDW and SS phase is accompanied by the softening of both hole-like excitations (blue line) on the even sites and particle-like excitations (red line) on the odd sites in the CDW phase. In the SS phase, the excitations form either a gapless mode (blue line),  also present in the SF phase, or a gapped mode (red line).

We further infer the nature of these excitations from a coherent state analysis at fixed and small momentum \cite{Huber2007}, see also Appendix C. We observe that the gapless mode is mainly constituted of the modulation of the phase of the superfluid order parameters. The nature of the gapped mode depends on the strength of the tunneling. In the vicinity of the CDW to SS phase transition, it is characterized by a large modulation of the amplitude of the superfluid order parameters but negligible modulation of density imbalance and phase of the superfluid order parameters. We thus  associate this excitation with an amplitude-Higgs mode \cite{Endres2012} near the CDW to SS phase transition. Approaching the SS to SF phase transition, the character of this excitation mode changes smoothly and now shows a much stronger imbalance modulation but a weak modulation of the amplitude of the superfluid order parameters. In this regime, we associate this excitation mode with a roton mode \cite{Mottl2012a} which is a hallmark of long-range interacting systems. 

In summary, we studied a system with competing global- and short-range interactions in the context of an extended Bose-Hubbard model and analyzed the nature of the low energy excitations of the system. Our findings are experimentally accessible since the cavity output field provides real-time access to both the imbalance order parameter and to the excitation spectra \cite{Landig2015}.

\appendix
\section{APPENDIX A: COMPARISON OF INFINITE-RANGE AND NEAREST-NEIGHBOUR INTERACTIONS}
In the following, we show on a mean-field level the formal equivalence of the Bose-Hubbard model including infinite-range interactions and the Bose-Hubbard model including nearest neighbour interactions. Combining the approximation made in equation (\ref{ThetaApprox}) with Hamiltonian (\ref{BHHam}) describing the system with infinite-range interactions results in the effective Hamiltonian:

\begin{equation}
\begin{split}
\hat{H}' =  - t \sum_{\langle i,j \rangle}\left(\hat{b}_i^\dag \hat{b}_{j} + \mathrm{h}.\mathrm{c}.\right) + \frac{U_s}{2} \sum_{i\in e,o} \hat{n}_{i}(\hat{n}_{i}-1) \\
   - U_l \theta \left(\sum_{i\in e}{\hat{n}_{i}} - \sum_{i\in o}{\hat{n}_{i}}\right) + U_l\frac{K}{4}\theta^2 - \mu\sum_{i\in e,o}{\hat{n}_{i}}\,.
\end{split}
\label{BHHam1}
\end{equation}

The Bose-Hubbard Hamiltonian with nearest neighbor interactions of strength $U_{nn}$ reads
\begin{equation}
\begin{split}
\hat{H}_{nn} =  - t \sum_{\langle i,j \rangle}\left(\hat{b}_i^\dag \hat{b}_{j} + \mathrm{h}.\mathrm{c}.\right) + \frac{U_s}{2} \sum_{i\in e,o} \hat{n}_{i}(\hat{n}_{i}-1) \\
+ U_{nn}\sum_{\langle i,j \rangle} \hat{n}_i\hat{n}_j - \mu\sum_{i\in e,o}{\hat{n}_{i}}\,.
\end{split}
\label{BHHam2}
\end{equation}
To show the equivalence with the above effective Hamiltonian (\ref{BHHam1}), we rewrite the nearest neighbor interaction term as
\begin{equation*}
\begin{split}
\sum_{\langle i,j \rangle} \hat{n}_i\hat{n}_j = \frac{z}{2}\sum_{i}\hat{n}_{i}(\hat{n}_{i}-1) + \frac{z}{2}\sum_{i} \hat{n}_{i} \\
- \frac{1}{2}\sum_{\langle i,j \rangle}(\hat{n}_{i}-\hat{n}_{j})^2
\end{split}\,,
\end{equation*}
where $z$ denotes the coordination number. Repulsive nearest-neighbor interactions  favor SS and CDW density waves similar to our description \cite{Goral2002}. This suggests to introduce also here an order parameter $\theta = \langle \hat{n}_i\rangle - \langle \hat{n}_j\rangle$, which then allows us to simplify the last term in mean-field approximation to:

\begin{align*}
\sum_{\langle i,j \rangle}(\hat{n}_{i}-\hat{n}_{j})^2 & \approx \sum_{\langle i,j \rangle} \Big[2(\hat{n}_{i}-\hat{n}_{j})\langle (\hat{n}_{i}-\hat{n}_{j}) \rangle - \langle (\hat{n}_{i}-\hat{n}_{j})\rangle^2 \Big]\\
& = 2z \theta \left( \sum_{i \in e} n_i - \sum_{i \in o} n_i \right) - \frac{z}{2} K \theta^2
\label{BHHam4}
\end{align*}

Hamiltonian \eqref{BHHam2} can then be expressed as
\begin{equation}
  \begin{split}
\hat{H}'_{nn} =  - t \sum_{\langle i,j \rangle}\left(\hat{b}_i^\dag \hat{b}_{j} + \mathrm{h}.\mathrm{c}.\right) + \frac{U_s-zU_{nn}}{2} \sum_{i\in e,o} \hat{n}_{i}(\hat{n}_{i}-1)
\\ - U_{nn} \theta \left( \sum_{i \in e} n_i - \sum_{i \in o} n_i \right) +\frac{z K}{4} U_{nn}\theta^2 - (\mu+\frac{z}{2} U_{nn})\sum_{i\in e,o}{\hat{n}_{i}}\,.
\end{split}
\end{equation}

Comparing $\hat{H}'_{nn}$ and $\hat{H}'$ shows the equivalence between global-range interactions and nearest neighbor interactions on the mean-field level, up to a renormalization of chemical potential and short range interactions.

\section{APPENDIX B: VALIDITY OF THE EFFECTIVE HAMILTONIAN}

Hamiltonian Eq.\eqref{BHHam} is obtained using the lowest-band approximation, valid for not too shallow lattices, where the energy gap to the first excited band is large compared to the energy scale of the short-range and the global-range interaction. Since both the energy gap between the two lowest bands as well as the short-range interaction strength $U_s$ scale for deeper lattices with $\sqrt{V}$ (where $V$ is the square lattice depth along $x-$ and $z-$direction), the short-range interaction strength will never exceed the band gap for increasing $V$. In contrast, the global-range interaction strength $U_l$ scales as $V e^{-1/\sqrt{V}}$ and additionally depends on the detuning $\Delta_\mathrm{c}$. Very large values for $U_l$ can thus require the inclusion of higher bands in the description. However, for the parameter regime considered in this work (and also for the parameter regime experimentally explored in \cite{Landig2016} near the phase transitions), we estimate the influence of higher bands to be marginal.

Light scattered into the resonator will give rise to a lattice potential $V_{\rm{cav}}\propto \cos^2(k x)$ acting on the atoms, which we neglect in our calculations. Its strength is
\[
V_{\rm{cav}} \approx U_0\frac{M_0^2\eta^2}{\Delta_c^2 + \kappa^2}\langle\hat{\Theta}\rangle^2\,,
\]
where $U_0$ is the maximum dispersive shift per atom. $V_{\rm{cav}}$ scales quadratically with the total imbalance $\langle\hat{\Theta}\rangle$ which itself is linearly depending on system size and density.  It has the same scaling with lattice depth $V$ as the global-range interactions. Again, for the parameter regime considered in this work (and mostly for the parameter regime experimentally explored in \cite{Landig2016}), we estimate the influence of this additional lattice onto the phase transitions to be negligible, see also \cite{Chen2016a}.

The additional dynamic lattice $V_{\rm{cav}}$ can however also be made to influence the phase diagram strongly by imbalancing the two applied lattices along the $x-$ and $z-$directions. Experimentally, this can be used to broaden the highly correlated SS regime near the transition to the CDW phase. In the limiting case of a vanishing depth of the externally applied lattice along the $x-$direction, the MI phase disappears and the dynamic lattices are solely responsible for reaching the CDW phase \cite{Bakhtiari2015, Klinder2015a}.

\section{APPENDIX C: COHERENT STATE ANALYSIS IN THE SS REGION}
A coherent state analysis \cite{Huber2007} is used to analyze the nature of the excitations in the SS phase. We diagonalize the effective Hamiltonian $\hat{H}_\mathrm{exc}$ and obtain operators generating the low-energy excitations in the system, ${\hat{\beta}_{i,q}}^{\dagger}$ at quasi-momentum $q$, where $i \in \{1,2\}$ refers to a particular excitation. We consider coherent states $|B_{i,q}\rangle$, which are the eigenstates of the corresponding annihilation operators with amplitude $B_{i,q}$. Depending on the parity of the site index $j$ being even ($e$) or odd ($o$), we define the following expectation values:
\[
\delta n^{i,q}_j = \langle B_{i,q} |\hat{n}_j| B_{i,q}\rangle - n_{e/o}
\]
\[
\delta \theta^{i,q}_{j'} = \langle B_{i,q} |\hat{n}_{j'+\frac{1}{2}}-\hat{n}_{j'-\frac{1}{2}}| B_{i,q}\rangle - \theta
\]
\[
\psi^{i,q}_j = \langle B_{i,q} |\hat{b}_j| B_{i,q}\rangle
\]
\[
\delta |\psi^{i,q}_j| = |\psi^{i,q}_j| - \psi_{e/o}
\]
\[
\arg(\psi^{i,q}_j) = \arctan\left( {\frac{\rm{Im}(\psi^{i,q}_j)}{\rm{Re}(\psi^{i,q}_j)}} \right)
\]
Here, $\delta n^{i,q}_j$ is the density variation on site $j$ from the average value $n_{e/o}$ on even respectively odd sites. $\delta \theta^{i,q}_{j'}$ describes the imbalance variation from the mean value $\theta$. The position of a pair of sites is labeled $j'= k + \frac{1}{2}$ with $k$ being an odd number. $\psi^{i,q}_j$ is the superfluid order parameter on site $j$ in presence of the excitation $|B_{i,q}\rangle$ and is defining the change $\delta |\psi^{i,q}_j|$ in amplitude of the superfluid order parameter from its mean value $\psi_{e/o}$ on even respectively odd sites. Finally, $\arg(\psi^{i,q}_j)$ is the phase of the superfluid order parameter as a function of position. 

\begin{figure}[h]
\includegraphics[width=1.0\columnwidth]{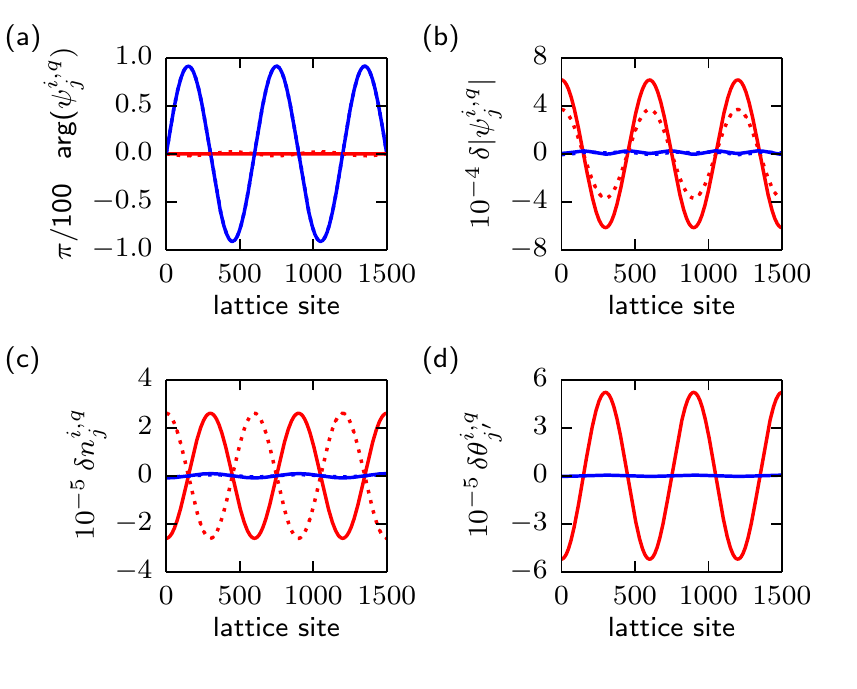}
\caption{Expectation values in the SS state in presence of an excitation $|B_{i,q=1/600}\rangle$. Shown is the variation around its mean value for (a) the phase of the superfluid order parameter, (b) the amplitude of the superfluid order parameter, (c) the density and (d) the imbalance. Excitation $i=1$ ($i=2$) is displayed in blue (red). Expectation values on even (odd) sites are marked as solid (dashed) line.  Data is shown for $\rho = 1/2$, $\tilde{U_l} = 0.3$ and $\tilde{t} =  0.2314$.}\label{appendix_fig1}
\end{figure}

From the response of the different expectation values upon an excitation, as shown in Fig.~\ref{appendix_fig1}, we infer about the nature of these excitations. We associate excitation mode $i=1$ with a gapless sound mode, since it is almost purely associated with the modulation of the argument $\arg(\psi^{i,q}_j)$ of the superfluid order parameter and is in phase for even and odd sites. This excitation is shown as blue symbols in Fig.~\ref{fig4} (b).  In contrast, excitation mode $i=2$ (displayed by red symbols in  Fig.~\ref{fig4} (b)) is mostly associated with the modulation of the amplitude of the superfluid order parameter. The modulation strength is different on even and odd sites, but they are in phase with respect to each other. For this mode, there is also an out-of-phase modulation of the particle density with equal strength on even and odd sites, which finally gives rise to an imbalance modulation. 

\begin{figure}[h]
\includegraphics{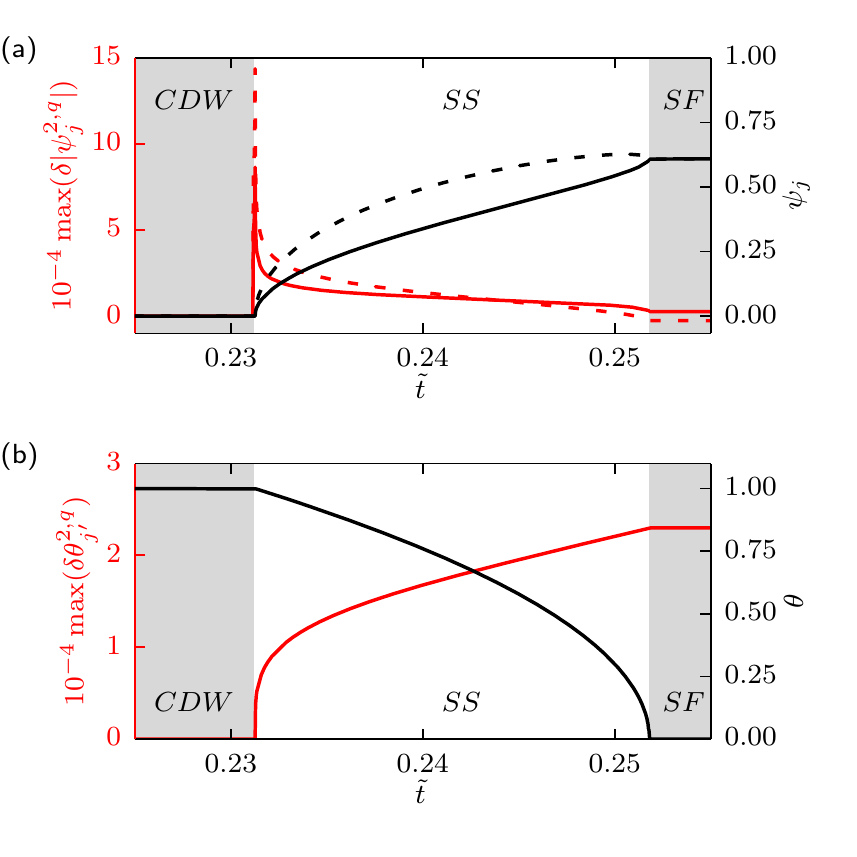}
\caption{
Modulation amplitudes for the $i=2$ excitation in the SS phase of (a) the variation of the amplitude of the superfluid order parameter $\delta |\psi^{2,q}_j|$, and (b) the density imbalance $\delta \theta^{2,q}_{j'}$   as a function of $\tilde{t}$. The result for even (odd) lattice sites is shown as solid (dashed) line. The relative sign in panel (a) indicates whether the excitations are in-phase or out-of-phase. For comparison, also the mean values of the superfluid order parameter (panel a, right axis) respectively of the imbalance (panel b, right axis) are displayed, see also Figure 1b. Data is presented for $\rho = 1/2$ and $\tilde{U_l} = 0.3$ as a function of $\tilde{t}$ in the presence of a fixed excitation $B_{2,q=1/600}$.}
\label{appendix_fig2}
\end{figure}

To fully characterize the excitation mode $i=2$, we inspect the modulation amplitudes of $\delta |\psi^{2,q}_j|$ and $\delta \theta^{2,q}_{j'}$ in the SS phase as a function of $\tilde{t}$, shown in Fig.~\ref{appendix_fig2} together with the mean values of the order parameters. The modulation amplitude of $\delta |\psi^{2,q}_j|$ is largest close to the CDW-to-SS phase transition, and decreases towards the SS-to-SF phase transition. The imbalance modulation amplitude shows the opposite behavior and is largest close to the SS-to-SF phase transition. We thus interpret excitation mode $i=2$ close to the CDW-to-SS transition as an amplitude-Higgs mode \cite{Huber2007} that changes its character to a roton-type mode \cite{Mottl2012a} close to the SS-to-SF phase transition. 

\vspace{1cm}

\begin{acknowledgments}
\emph{Acknowledgements:} We thank T. Esslinger, L. Hruby, R. Landig, M. Landini, R. Mottl, and G. Morigi for stimulating discussions. Financial funding from Synthetic Quantum Many-Body Systems (European Research Council advanced grant), the EU Collaborative Project TherMiQ (Grant Agreement 618074), SBFI support for Horizon2020 project QUIC, and SNF support including the NCCR QSIT and the DACH project 'Quantum Crystals of Matter and Light'. 
\end{acknowledgments}

\bibliographystyle{apsrev4-1}

\end{document}